\documentclass[
  journal=largetwo,
  manuscript=article-type,
  year=2020,
  volume=37,
]{cup-journal}

\usepackage{xcolor}
\usepackage{verbatim}

\usepackage{amsmath}
\usepackage{booktabs}

\usepackage{multicol}
\usepackage{multirow}

\newcommand{\jetcaf}{{\fontfamily{qcr}\selectfont JeTCAF}}
\newcommand{\relxil}{{\fontfamily{qcr}\selectfont relxillCp}}
\newcommand{\pexrav}{{\fontfamily{qcr}\selectfont pexrav}}
\newcommand{\tcaf}{{\fontfamily{qcr}\selectfont TCAF}}

\newcommand{\tbabs}{{\fontfamily{qcr}\selectfont TBABS}}
\newcommand{\ztbabs}{{\fontfamily{qcr}\selectfont ZTBABS}}

\newcommand{\polcon}{{\fontfamily{qcr}\selectfont POLCONST}}
\newcommand{\constant}{{\fontfamily{qcr}\selectfont CONST}}

\newcommand{\cutoffpl}{{\fontfamily{qcr}\selectfont CUTOFFPL}}
\newcommand{\gauss}{{\fontfamily{qcr}\selectfont GAUSS}}

\title{Spectro-polarimetric study to constrain accretion-ejection properties of MCG-5-23-16 using {\it IXPE} and {\it NuSTAR} observations}

\author{Santanu Mondal}
\affiliation{Indian Institute of Astrophysics, II Block, Koramangala, Bengaluru 560034, India}
\email[S. Mondal]{santanumondal.work@gmail.com, santanu.mondal@iiap.res.in}

\author{Rwitika Chatterjee}
\affiliation{Space Astronomy Group, ISITE Campus, U. R. Rao Satellite Center, ISRO, Bengaluru, 560037, India}

\author{Vivek K. Agrawal}
\affiliation{Space Astronomy Group, ISITE Campus, U. R. Rao Satellite Center, ISRO, Bengaluru, 560037, India}

\author{Anuj Nandi}
\affiliation{Space Astronomy Group, ISITE Campus, U. R. Rao Satellite Center, ISRO, Bengaluru, 560037, India}

\keywords{accretion disk, galaxies:active, galaxies: Seyfert, polarization, X-rays: individual: MCG-5-23-16} 

\begin{document}

\begin{abstract}
	We conducted a study on the X-ray polarization properties of MCG-5-23-16 by analyzing long-term monitoring data from {\it NuSTAR} jointly with {\it IXPE} observations made in May and November 2022. The re-analysis of {\it IXPE} data gives model-dependent polarization degree, PD (\%) = $1.08\pm0.66$ in the energy band $2-8$ keV, which agrees with previous studies within error bars. The model-independent analysis of PD poses an upper limit of $\leq3.8$ ($1\sigma$ level) for the same energy band. The observed upper limit of PD, along with broadband spectral analysis ($2-79$ keV) using an accretion-ejection based model, allowed us to derive the corona geometry (i.e. radius and height) and the accretion disk inclination ($\sim 33^\circ$).
Additional {\it NuSTAR} observations were also analyzed to gain insights into the accretion flow properties of the source and to estimate the expected polarization during those epochs with PD $\sim 4.3\%$. The radius and height of the corona varies between $28.2\pm3.1 - 39.8\pm4.6$ r$_s$ and $14.3\pm1.7-21.4\pm1.9$ r$_s$ respectively, with a mass outflow rate from the corona measuring $0.14\pm0.03-0.2\pm0.03$ Eddington rate ($\dot m_{\rm Edd}$). The estimated PD values were nearly constant up to a certain radial distance and height of the corona and then decreased for increasing corona geometry. The spectral analysis further provided an estimate for the mass of the central black hole $\sim 2\times 10^7$ M$_\odot$ and the velocity of the outflowing gas $\sim 0.16-0.19c$. A comparative broadband spectral study using reflection-based models estimates the disk inclination between $\sim 31^\circ\pm8^\circ-45^\circ\pm7^\circ$, and yields an expected PD of 3.4-6.0\%. We also found a weak reflection fraction and a less ionized distant reflecting medium. The expected PD measured using accretion-ejection and reflection models is less compared to the expected PD measured for a given disk inclination of $45^\circ$.
Our modeling of the disk-corona-outflows and polarization connection can be extended and validated with data from the recently launched \textit{XPoSat}, India's first X-ray Polarimeter Satellite, offering potential applications to other sources.
\end{abstract}

\section{Introduction}
\label{introduction}
Recent launches of {\it Imaging X-ray Polarimetry Explorer} \autocite[IXPE;][]{WeisskopfEtal2016SPIE.9905E..17W} and {\it X-ray Polarimeter Satellite} (XPoSat)\endnote{https://www.isro.gov.in/XPoSat\_X-Ray\_Polarimetry\_Mission.html}  Missions have renewed interest in measuring X-ray polarization from diverse astrophysical systems, in particular from accreting black hole (BH) systems of all scales. Accreting matter puffs up at the inner region of the disk, which not only upscatters soft photons through inverse Comptonization \autocite[][and references therein]{SunyaevTitar80,HaardtMaraschi1993,ChakrabartiTitarchuk1995,Done2007,MondalChak2013MNRAS.431.2716M,IyerEtal2015ApJ...807..108I} but also polarizes the radiation depending on its thermodynamic properties, geometry, and inclination \autocite{ConnorsEtal1980ApJ...235..224C,SunyaevTitar85}. Therefore, the measurement of polarization angle (PA) and degree (PD) depends on both corona (Compton cloud) properties as well as the photon energy. Additionally, the presence of outflows from the disk can also produce significant polarization due to scattering effects \autocite{BegelmaMcKee1983ApJ...271...89B}. The dependence of PD and PA on the optical depth and disk inclination was studied using different geometry of the corona in detail \autocite[][and references therein]{DovciakEtal2008MNRAS.391...32D,LiEtal2009ApJ...691..847L,SchnittmanKrolik2010ApJ...712..908S}. Therefore, if the corona properties of a system is well constrained, the PD can be estimated or the vice-versa. Hence, the spectro-polarimetric studies using broadband data may shed more light on the measurement of polarization and the environment of the corona around BHs.

MCG-5-23-16 is a Seyfert 1.9 galaxy \autocite{Veronetal1980} located at redshift z = 0.00849 \autocite{WegnerEtal2003AJ....126.2268W}. The source is relatively bright in X-rays with F$_{2-10}$ = $7-10\times10^{-11}$ erg cm$^{-2}$ s$^{-1}$ \autocite{MattsonWeaver2004ApJ...601..771M} and characterized by moderate neutral absorption ($N_H \sim 10^{22}$ cm$^{-2}$). It has been extensively studied in the X-ray band to estimate corona properties and spectral high-energy cut-off using {\it INTEGRAL} and {\it NuSTAR} observations \autocite{BeckmannEtal2008A&A...492...93B,MolinaEtal2013MNRAS.433.1687M,BalokovicEtal2015ApJ...800...62B,ZoghbiEtal2017ApJ...836....2Z}. 

Furthermore, the X-ray spectrum of this source showed the presence of a soft excess and complex Fe K$\alpha$ emission with broad and asymmetric narrow line features in {\it ASCA} \autocite[][]{WeaverEtal1997ApJ...474..675W}, {\it BeppoSAX}, {\it Chandra}, and {\it XMM-Newton} \autocite{DewanganEtal2003ApJ...592...52D,BalestraEtal2004A&A...415..437B,BraitoEtal2007ApJ...670..978B,LiuEtal2024ApJ...963...38L} observations. It indicates the presence of two reflectors, one for a narrow core at 6.4 keV and the other for the broad component, possibly originating closer to the BH. Later, the broad line was explored with {\it Suzaku} \autocite{ReevesEtal2007PASJ...59S.301R} and reported that the inner radius of the disk is 20 r$_s$ that originated the line, where r$_s$ = $2GM_{BH}/c^2$ is the Schwarzschild radius, with an inclination $50^\circ$. Additionally, an absorption feature at 7.7 keV was unveiled in these data, pointing to the possible presence of ionized iron outflowing at $\sim 0.1c$. Recently, \cite{SerafinelliEatal2023MNRAS.526.3540S} estimated the disk inclination of 41$^\circ$ from the broad Fe K$\alpha$ line study using {\it NuSTAR} observations. \cite{MarinucciEtal2022MNRAS.516.5907M} estimated the disk inclination ${48^\circ}^{+12}_{-8}$ from broad Fe K$\alpha$ line profile. The disk inclination also found to lie between $40^\circ-50^\circ$ \autocite[see also][]{WeaverReynolds1998ApJ...503L..39W}. The mass of the central supermassive BH (SMBH) is $2\times10^7$ M$_\odot$ derived independently from both X-ray variability \autocite{PontiEtal2012A&A...542A..83P}  and infrared lines \autocite{Onorietal2017}.

Along with spectral and timing studies, X-ray polarization provides an independent tool to constrain the accretion properties and coronal geometry. It has been reported that the polarization is extremely sensitive to the geometry of the radiation emitting region and the photon field \autocite[][and references therein]{SchnittmanKrolik2010ApJ...712..908S,BeheshtipourEtal2017ApJ...850...14B,TamborraEtal2018A&A...619A.105T}. To constrain the source coronal geometry, MCG-5-23-16 was targeted by two pointings in 2022 with IXPE.

The joint analysis of {\it XMM-Newton}, {\it NuSTAR}, and {\it IXPE} observations in May 2022 by \cite{MarinucciEtal2022MNRAS.516.5907M} obtained a 4.7\% (at 99\% confidence level) upper limit for the polarization degree (PD) in the 2–8 keV energy band and ruled out the slab geometry of the corona. Authors also found a hint of alignment between the polarization angle and the accretion disk spin axis. Later, \cite{TagliacozzoEtal2023MNRAS.525.4735T} analyzed the {\it IXPE} data from both May and November 2022 epochs in coordination with {\it NuSTAR} and estimated the upper limit of PD to be 3.2 \% (at 99\% confidence level for one parameter of interest). Their preferred polarization angle in the $\sim 50^\circ$ direction may hint that the polarization of the primary emission is aligned with the Narrow Line Region, which was observed at $\sim 40^\circ$ position angle in Hubble Space Telescope's WFPC2 images \autocite{FerruitEtal2000ApJS..128..139F} and therefore, parallel to the accretion disk axis, similar to what was found in NGC\,4151 \autocite{GianolliEtal2023MNRAS.523.4468G}. Such alignments may contribute to the polarization measurement just below detection. The authors further compared their observed estimations with Monte Carlo simulations of the expected polarization properties for different geometries of the corona and disfavored the lamppost and cone-shaped corona.

Therefore, we see that polarization can be used as a probe to understand the geometry of the corona and accretion-ejection behavior around the central SMBH. The continuum models applied to study the spectral properties of the source MCG-5-23-16 mostly take into account radiative transfer processes. However, to date, no physically motivated accretion-ejection based models have been directly employed to fit and understand the polarization properties with the corona geometry or accretion flow parameters. In this work, we have re-analyzed the {\it IXPE} observations to confirm the previous studies jointly with {\it NuSTAR} data for both epochs (May and November 2022). Further, other data sets available in the {\it NuSTAR} archive were also analyzed to constrain the corona properties and to predict the polarization behavior. Moreover, we have also estimated the mass outflow rate for all epochs, that connect the disk-corona-outflow geometry with polarization from a single framework.    

The paper is organized as follows: in \S 2, we discuss the observation and data reduction procedure. In \S 3, we describe the spectro-polarimetric
data analysis results, and finally, we draw our conclusions in \S 4.

\section{Observation and Data Reduction}
Over the past decade, MCG-5-23-16 was observed by {\it NuSTAR} \autocite{Harrisonetal2013}, spanning from 2012 to 2022. Most recently, the source was observed in May and November 2022 by {\it IXPE} \autocite{WeisskopfEtal2016SPIE.9905E..17W}, launched on 2021 December 9, consisting of three Wolter-I telescopes, with 3 units of polarization-sensitive imaging X-ray detector units (DUs) at the respective foci.
We used all archival data from both observatories for this source and analyzed them. The observation log is given in Table \ref{table:observation}.

{\it IXPE} provides Level-2 data, which we cleaned and calibrated using standard \texttt{FTOOLS} tasks with the latest calibration files (CALDB 20230526) available in {\it IXPE} database. A circular region centered at the source of radius $60^{\prime\prime}$ \autocite[see][for details]{TagliacozzoEtal2023MNRAS.525.4735T} is used for all detectors. A source-free background region of radius $100^{\prime\prime}$ is used to produce background subtracted spectra for I, Q, and U Stokes for all three DUs. Additionally, we applied a weighting scheme, \texttt{STOKES=NEFF} \autocite{DiMarcoEtal2022AJ....163..170D} in \texttt{XSELECT} to generate the final Stokes spectra in the $2-8$ keV energy band.

The {\it NuSTAR} data were extracted using the standard 
{\sc NUSTARDAS v1.3.1}
\endnote{\url{https://heasarc.gsfc.nasa.gov/docs/nustar/analysis/}} software. We 
ran a {\sc nupipeline} task to produce cleaned event lists and 
{\sc nuproducts} along with CALDB version 20231121 to generate the spectra. We used a region of 
$25^{\prime\prime}$  for the source and $40^{\prime\prime}$ for the background 
using {\sc ds9} \autocite{JoyeDS92003ASPC..295..489J}. The data were grouped with a 
minimum of 30 counts in each bin using {\it grppha} command. For the analysis of each epoch of 
observation, we used the data of both {\it IXPE} and {\it NuSTAR} in the energy range of $2 - 8$ keV and $3 - 79$ keV. We used {\sc XSPEC}\endnote{\url{https://heasarc.gsfc.nasa.gov/xanadu/xspec/}} \autocite{Arnaud1996} version 12.12.0 for spectral analysis. The hydrogen column density for Galactic absorption set fixed to $7.8\times 10^{20}$cm$^{-2}$ \autocite{HI4PICollabor2016A&A...594A.116H,Kalberlaetal2005} during the spectral fitting throughout.

\begin{table}
\centering
\caption{\label{table:observation} Log of observations of MCG-5-23-16. The $^\ast$ denotes data used for joint spectro-polarimetric analysis.}
\begin{tabular}{ccccccc}
\hline
  Obs. Id. &Exp. & Epoch  & Start  &End   \\
           &(ks) &        & Date   &Date    \\
\hline
&&{\it IXPE}&&\\
\hline
01003399&436&I1&2022-05-14&2022-05-31  \\
02003299&641&I2&2022-11-06&2022-11-23  \\
\hline
           &&{\it NuSTAR}&&\\
\hline
60701014002$^\ast$&84&N1&2022-05-21&2022-05-23 \\
90801630002$^\ast$&86&N2&2022-11-11&2022-11-12 \\
60001046008&221&N3&2015-03-13&2015-03-18 \\
60001046006&98&N4&2015-02-21&2015-02-23 \\
60001046004&210&N5&2015-02-15&2015-02-20 \\
60001046002&160&N6&2013-06-03&2013-06-07 \\
10002019001&34&N7&2012-07-11&2012-07-11  \\
\hline
\end{tabular}
\end{table}

\begin{table}
\centering
\caption{\label{table:polparams} Results of model-independent polarimetric analysis in the $2-8$ keV band.}
\begin{tabular}{ccc}
\hline
Parameter & I1   &I2   \\ \hline
Q/I (\%)  & $0.27\pm 1.26$  & $-0.15\pm 1.16$\\
U/I (\%)  & $1.33\pm 1.26$  & $1.10\pm 1.16$\\
PD (\%)   & $1.36\pm 1.26$  & $1.11\pm 1.16$\\
PA (\%)   & $39.34\pm 26.56$  & $48.94\pm 29.79$\\
MDP$_{99}$ (\%)   & 3.83 & 3.50\\ 
\hline
\end{tabular}
\end{table}

\section{Data Analysis and Results}

\subsection{Model-independent and Model-dependent {\it IXPE} Analysis}
The processed Level-2 data were used for polarimetric analysis, using \texttt{IXPEOBSSIM} software v30.0.0 \autocite{BaldiniEtal2022SoftX..1901194B}. Source and background events were extracted using \texttt{XPSELECT} task. We obtained the polarization parameters in the entire {\it IXPE} band of $2-8$ keV using the model-independent PCUBE algorithm of the \texttt{XPBIN} task \autocite{KislatEtal2015APh....68...45K}. 
Although, there are reports in the literature on the estimation of polarization, here we have re-done the analysis for the sake of completeness.

\begin{figure*}
	\centering 
	\includegraphics[width=0.45\textwidth, angle=0]{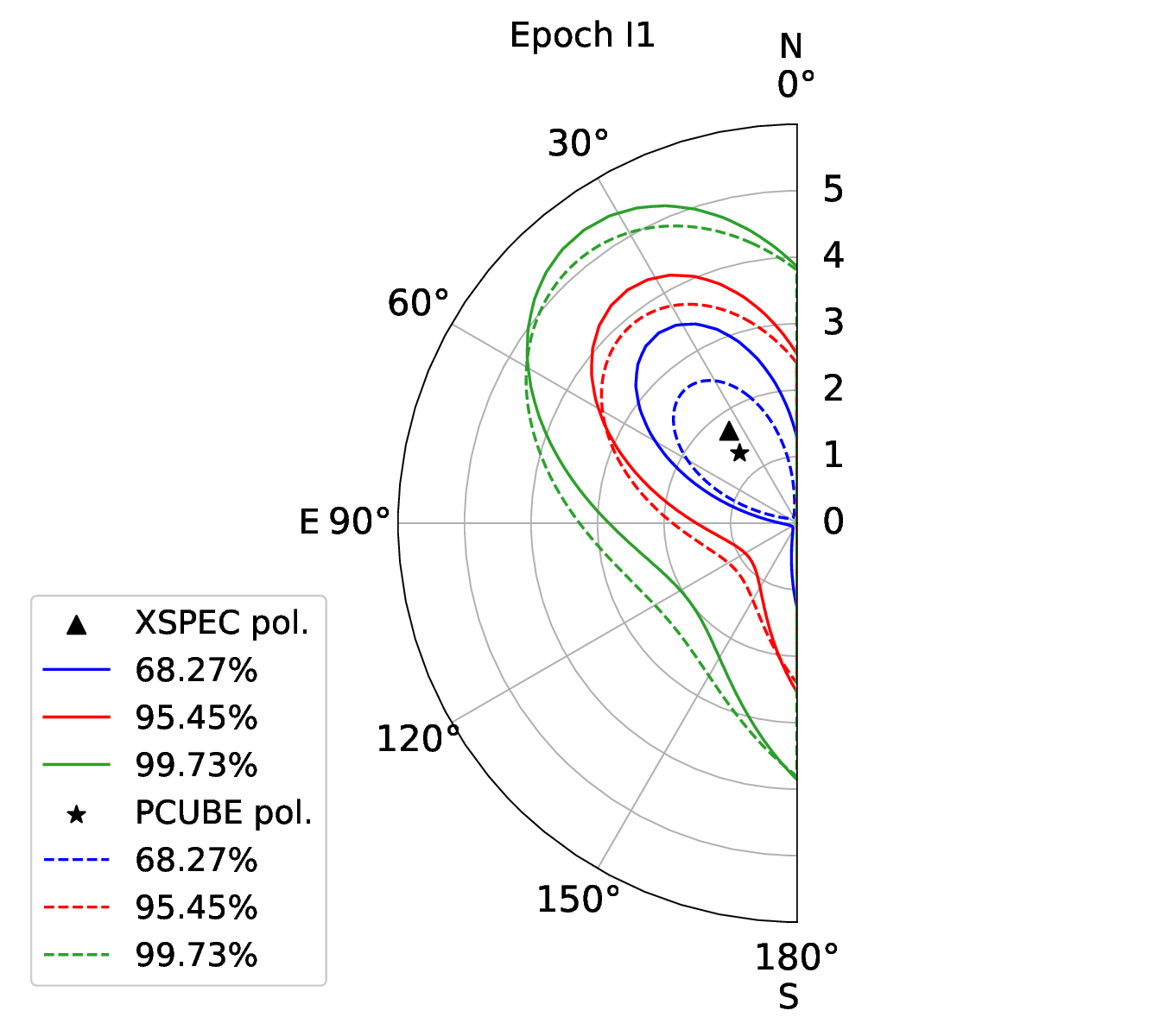}
    \includegraphics[width=0.45\textwidth, angle=0]{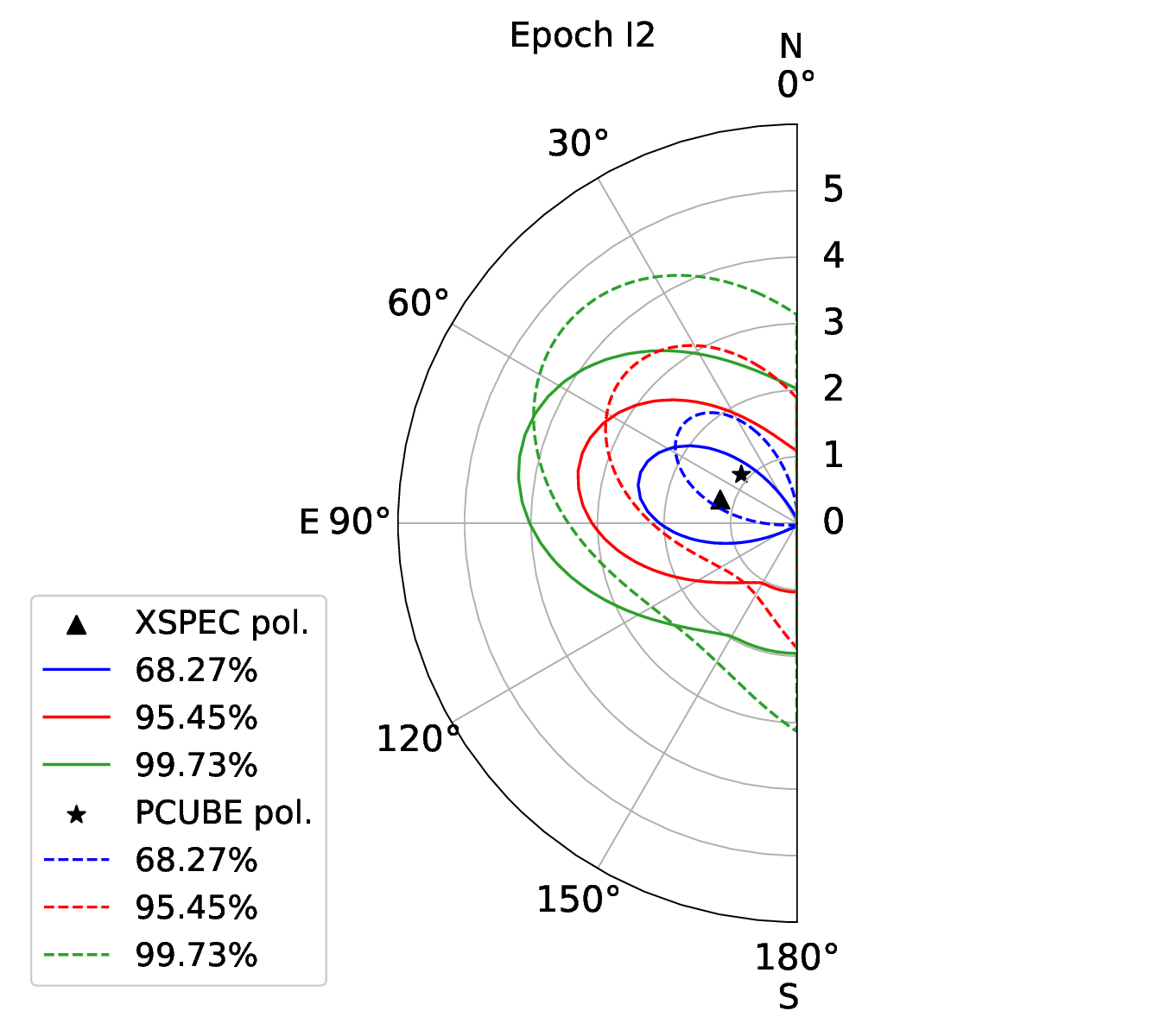}
	\caption{XSPEC (solid lines) and PCUBE (dashed lines) contour plots between polarization degree PD and angle PA for May (I1; left) and November (I2; right) epochs in 2022. The blue, red, and green contours denote $1\sigma$, $2 \sigma$, and $3\sigma$ confidence levels. See the text for details.} 
	\label{fig:pcube_contours}%
\end{figure*}

The results of our model-independent polarimetric analysis for the two epochs are summarized in Table \ref{table:polparams}. We obtain only upper limits of the PD for both epochs (below the minimum detectable polarization at the 99\% level).
The measured PD and PA in $2-8$ keV for the two epochs (I1 and I2) are shown in Fig. \ref{fig:pcube_contours} along with the contours. The dashed lines correspond to PCUBE contours with 1$\sigma$ (blue), 2$\sigma$ (red), 3$\sigma$ (green) levels. The PD upper limit (99\% confidence level) is $3.8\%$ and $3.5$\% in epochs I1 and I2 respectively. Our estimated upper limit of PD in I2 is consistent with the literature, however, in I1 is lower \citep{MarinucciEtal2022MNRAS.516.5907M,TagliacozzoEtal2023MNRAS.525.4735T}.

Further, we also performed a spectro-polarimetric 
model-dependent fit \citep{strohmayer2017} using a phenomenological model, consisting of an absorbed \cutoffpl\, convolved with a constant polarization \polcon\, which reads in \texttt{XSPEC} as \constant*\ztbabs*\polcon*\cutoffpl. The \cutoffpl\, model has two spectral parameters, viz. cutoff energy (E$_{\rm cut}$) and photon index ($\Gamma$). The Stokes spectra for all three DUs in the energy band $2-8$ keV are fitted simultaneously. A {\tt wilm} abundance \autocite{Wilmsetal2000} is used throughout the analysis. Since the determination of hydrogen column density may not be reliable owing to the lower energy limit of \textit{IXPE} or \textit{NuSTAR}, we freeze the value of intrinsic $N_H$ to $2.1\times 10^{22}$~cm$^{-2}$, which is the average value determined from \textit{Chandra} observations from 2000 to 2020 \autocite{LiuEtal2024ApJ...963...38L} and the Galactic $N_H$ fixed to $7.8\times 10^{20}$~cm$^{-2}$.

We began by leaving all the spectral and polarimetric parameters to vary freely. We found that the obtained values of PD and PA are highly sensitive to the values of the spectral parameters. Keeping in mind the limited energy band and the spectral sensitivity of \textit{IXPE}, we turned to simultaneous \textit{NuSTAR} observations for obtaining robust spectral fits. In addition to the \cutoffpl\, model, the {\it NuSTAR} data also required an emission feature at the Fe K line energy for a good fit. The full model reads in \texttt{XSPEC} as \ztbabs*(\gauss+\cutoffpl), which returned cutoff energy $\sim120$ keV, which is consistent with the previous findings by \autocite{MarinucciEtal2022MNRAS.516.5907M} and \autocite{TagliacozzoEtal2023MNRAS.525.4735T}.

Subsequently, we re-did the \textit{IXPE} fits by freezing the values of $E_{\mathrm{cut}}$ and $\Gamma$ to the best-fit values from the \textit{NuSTAR} fit of the respective epoch, which is expected to give a more robust estimation of PD and PA. Note that the addition of a Gaussian line and reflection component neither improves the quality of the fit nor affects the value of the polarisation parameters, hence we did not include these two components while fitting the \textit{IXPE} data. We obtain good fits for I1 and I2 using this model, with $\chi^2_{\mathrm{red}}=0.99$ and 0.92 respectively. We determine the PA = $36.3^\circ\pm19.4^\circ$ and $72.9^\circ\pm23.8^\circ$ and the PD (\%) = $1.72\pm1.08$ and $1.21\pm0.89$ for the epochs I1 and I2 respectively. However, since the polarization measurements have a low significance, their values as well as their apparent variation across the epochs should be interpreted with caution. The confidence contours from the model-dependent analysis are shown in Fig. \ref{fig:pcube_contours}. 
The model-dependent polarimetric results are summarized in Table \ref{table:PolMoDep}. Note that it was mentioned in \cite{TagliacozzoEtal2023MNRAS.525.4735T} that the model-dependent contours match with the PCUBE contours. However, for I2, we found a small difference between the PA obtained using the two approaches, and the possible reason for this is not clear yet.

\begin{table}
\centering
\caption{\label{table:PolMoDep} All Stokes spectra for all three DUs of {\it IXPE} are fitted using \ztbabs*\polcon*\cutoffpl\, model. The best-fitted parameters of the model-dependent polarimetric analysis are summarized here. All errors are estimated at the $1\sigma$ level. The $N_H$ value is frozen to $2.1\times 10^{22}$~cm$^{-2}$.}
\begin{tabular}{cccc}
\hline
 Parameters & I1 & I2 & (I1+I2)  \\
\hline
PD (\%)        &$1.72\pm1.08$&$1.21\pm0.89$ & $1.08\pm0.66$ \\
PA ($^\circ$) &$36.3\pm19.4$&$72.9\pm23.8$ &  $55.6\pm19.1$ \\
$\Gamma$          &$1.66$&$1.64$ & $1.98\pm0.02$ \\
$E_{\rm cut}$ (keV)&$122.2 $ &$124.2$ & 123  \\
$\chi^2/dof$      & 1312/1330  &1231/1330 & 2572/2659 \\
\hline
\end{tabular}
\end{table}

Further, since the PD measurements for both epochs have a low significance (from both model-dependent and model-independent approaches), we attempt to fit the two epochs simultaneously seeking a more significant polarization estimation. Since the E$_{\rm cut}$ and $\Gamma$ values obtained from the {\it NuSTAR} fits do not vary significantly, we tie these values across I1 and I2. Then, we perform  
the simultaneous fit by constraining them across the epochs. A constant multiplicative factor has been included in the model to account for the cross-calibration across DUs as well as flux changes across epochs. 
The simultaneously fitted PD is $\sim10-40$\% less than the individual epochs and the PA is nearly the mean of the individual analysis since it is done under the assumption that the polarization characteristics across the two epochs do not differ significantly. The model parameters are summarized in Table~\ref{table:PolMoDep}.

\subsection{Spectral Analysis of {\it IXPE} and {\it NuSTAR}}
We performed a broadband spectroscopic analysis combining the $2-8$ keV {\it IXPE} spectra and the $3-79$ keV {\it NuSTAR} spectra from May and November 2022. The broadband energy range ($2-79$ keV) covered by the spectra of MCG-5-23-16 showed a broad Fe K line and a Compton hump above 10 keV. A Gaussian component was required which was free to vary, however, the line energy fitted at $\sim 6.3$\,keV for all observations. 
 
As discussed earlier, the spectra showed the presence of another component that produces the excess emission, which can either be due to reflection \autocite{Done2007} or a signature of mass loss \citep{Chakrabarti1999} from the corona region. Therefore, we have used two different sets of models one is the accretion-ejection based \jetcaf\, model \citep{MondalChakrabarti2021}, and the others are reflection-based \pexrav\, \citep{MagdziarzZdzi1995MNRAS.273..837M} and \relxil\,models \citep{DauserEtal2014MNRAS.444L.100D,GarciaEtal2014ApJ...782...76G}. Additionally, we have analyzed all the {\it NuSTAR} archival data for the source MCG-5-23-16 covering a decade of observations (see Table \ref{table:observation}) to understand the variation of accretion and reflection parameters and predict the polarization degree from the model-fitted accretion disk parameters.

\subsubsection{Accretion-ejection based \jetcaf\, model}
\begin{figure}
    \centering
    \includegraphics[height=6.5cm]{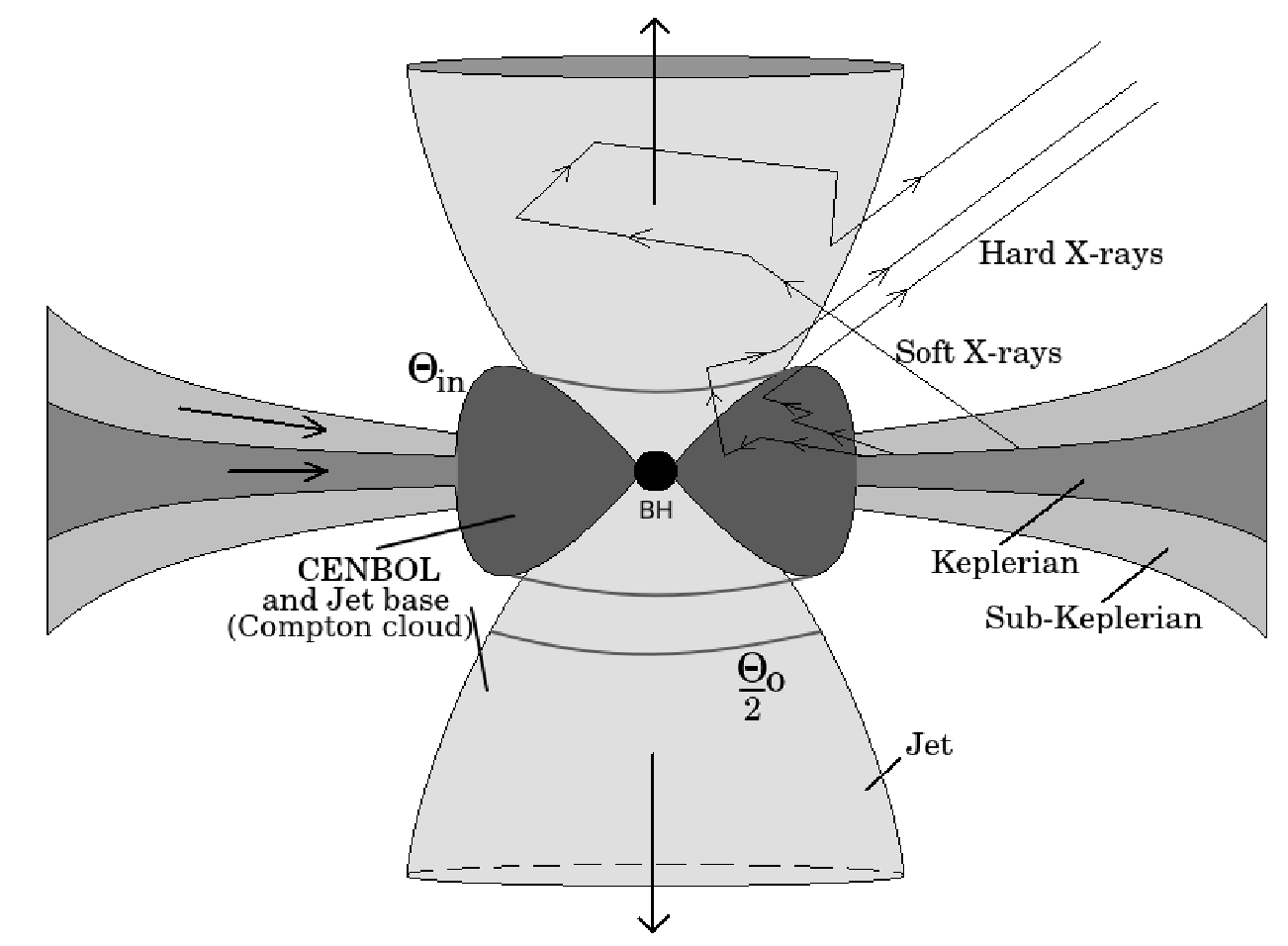}
    \caption{Illustration of the \jetcaf\, model. The Keplerian disk is residing at the equatorial plane while the sub-Keplerian halo component is above and below the equatorial plane, respectively. The deep gray inner region is the hot Compton cloud and the conical funnels above and below this region show the mass outflow or jet. The zig-zag arrows indicate the scattering of soft disk photons by different Comptonizing mediums. The figure is adapted from \cite{MondalChakrabarti2021}.}
    \label{fig:jetcaf-cartoon}
\end{figure}

As the accreting matter moves closer to the BH its velocity increases and becomes supersonic, which may or may not form shocks \autocite{Chakrabarti1989} depending on the satisfaction of the Rankine-Hugoniot conditions. The location of the formation of the shock is known as the boundary layer of the corona. The same shocked region also drives jets/outflows and behaves as the base of the jet \autocite{Chakrabarti1999}. Therefore, we further fit the data using an accretion-ejection-based jet in a two-component advective flow model \autocite[or \jetcaf,][]{MondalChakrabarti2021} to understand the accretion-ejection and polarization behavior of the source. The \jetcaf\,model is the updated version of \tcaf\,model \autocite{ChakrabartiTitarchuk1995} that takes into account the spectral signature of jet or mass outflows. The model with flow geometry is illustrated in Fig. \ref{fig:jetcaf-cartoon}. The \jetcaf\, model takes into account radiation mechanisms at the base of the jet and the bulk motion effect by the outflowing jet on the emitted spectra in addition to the Compton scattering of soft disk photons by the hot electron cloud inside the corona. The \jetcaf\,model has six parameters, namely (i) the mass of the BH ($M_{\rm BH}$), (ii) the Keplerian disk accretion rate ($\dot m_{\rm d}$), (iii) the sub-Keplerian halo accretion rate ($\dot m_{\rm h}$), (iv) the size of the dynamic corona or the location of the shock ($X_{\rm s}$ in $r_g=2GM_{\rm BH}/c^2$ unit), (v) the shock compression ratio ($R$), and (vi) the outflow/jet collimation factor ($f_{\rm col}$), i.e.,the ratio of the solid angle subtended by the outflow to the inflow ($\Theta_o/\Theta_{\rm in}$). 

In general, an increase in $\dot m_d$ makes the spectrum softer as the increased number of soft photons from the disk cools the corona and lowers the outflow rate \autocite{MondalEtal2014Ap&SS.353..223M}. While for increased $\dot m_h$ spectrum becomes harder as the corona is hotter and bigger, therefore, soft photons can gain more energy through scattering \citep{Chakrabarti1997,MondalChak2013MNRAS.431.2716M}. That also elevates the mass outflow rate. The increase in $X_s$ also makes the spectrum harder as more hard photons from the hot corona contribute to it. In this model, the $f_{col}$ parameter is incorporated to take into account the effect of outflows in addition to \tcaf\, parameters. A higher value of this parameter implies that the jet is less collimated due to a higher outflowing angle, which makes the spectrum harder.  
As the mass is a parameter, along with the accretion rates and corona parameters, the \jetcaf\,model has the potential to estimate the mass of the central SMBH as previously done for several BHs using \tcaf, if there is no robust estimation beforehand \autocite{IyerEtal2015ApJ...807..108I,MollaEtal2017ApJ...834...88M,Nandietal2019,MondalEtal2022A&A...662A..77M}. For the host galaxy absorption, we used the \ztbabs\, model. The full model reads in {\it XSPEC}\footnote{At present the model is not publicly available in {\it XSPEC}, however, a preliminary version of the FITS file is in the testing phase which can be available soon upon request.} for {\it IXPE} and {\it NuSTAR} as \tbabs*\ztbabs* (\jetcaf+\gauss). 

\begin{figure*}
\centering{
\hspace{-1.5cm}
\includegraphics[height=10.0truecm,width=7.0truecm,angle=270]{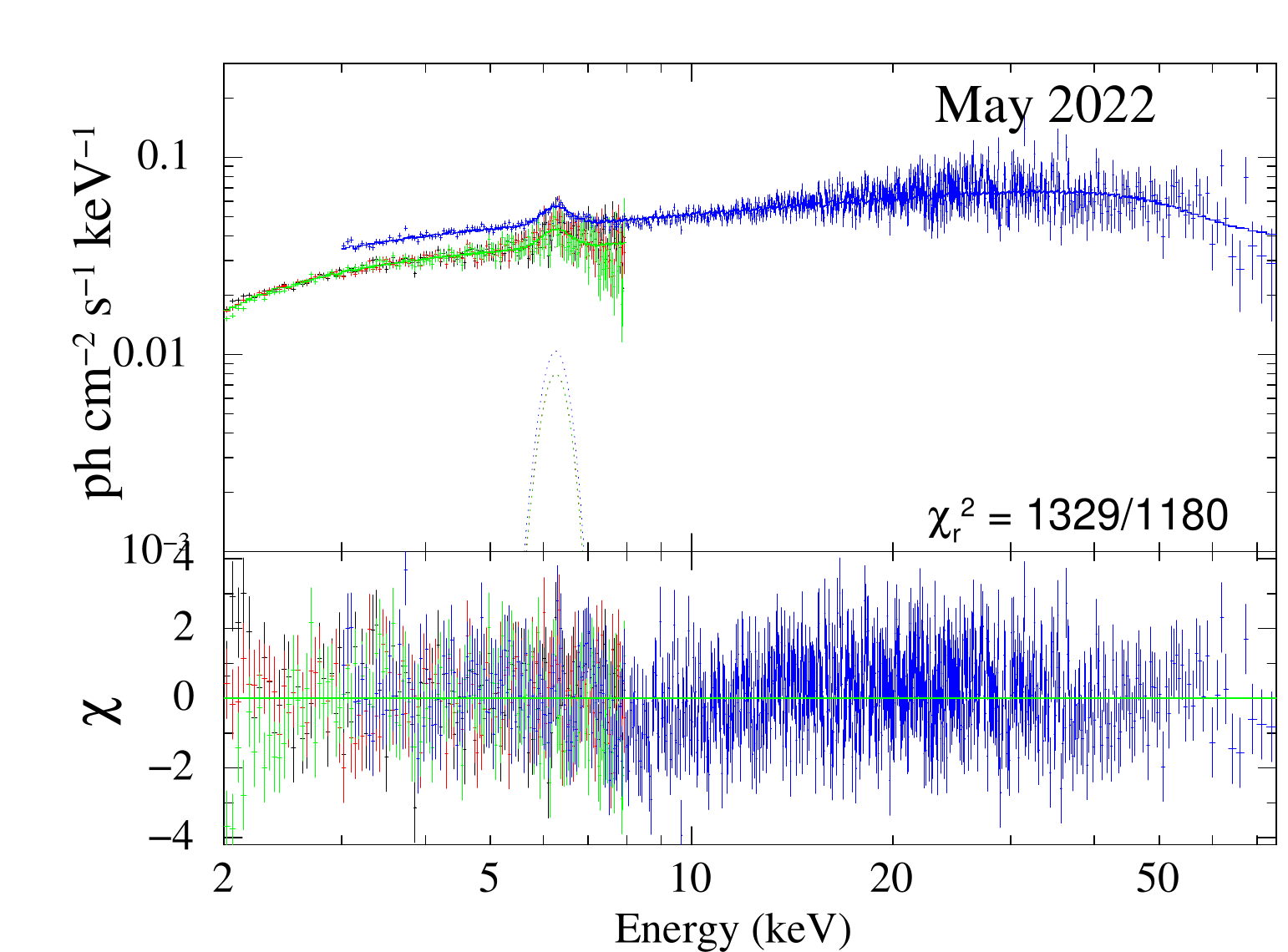}
\hspace{-0.7cm}
\includegraphics[height=10.0truecm,width=7.0truecm,angle=270]{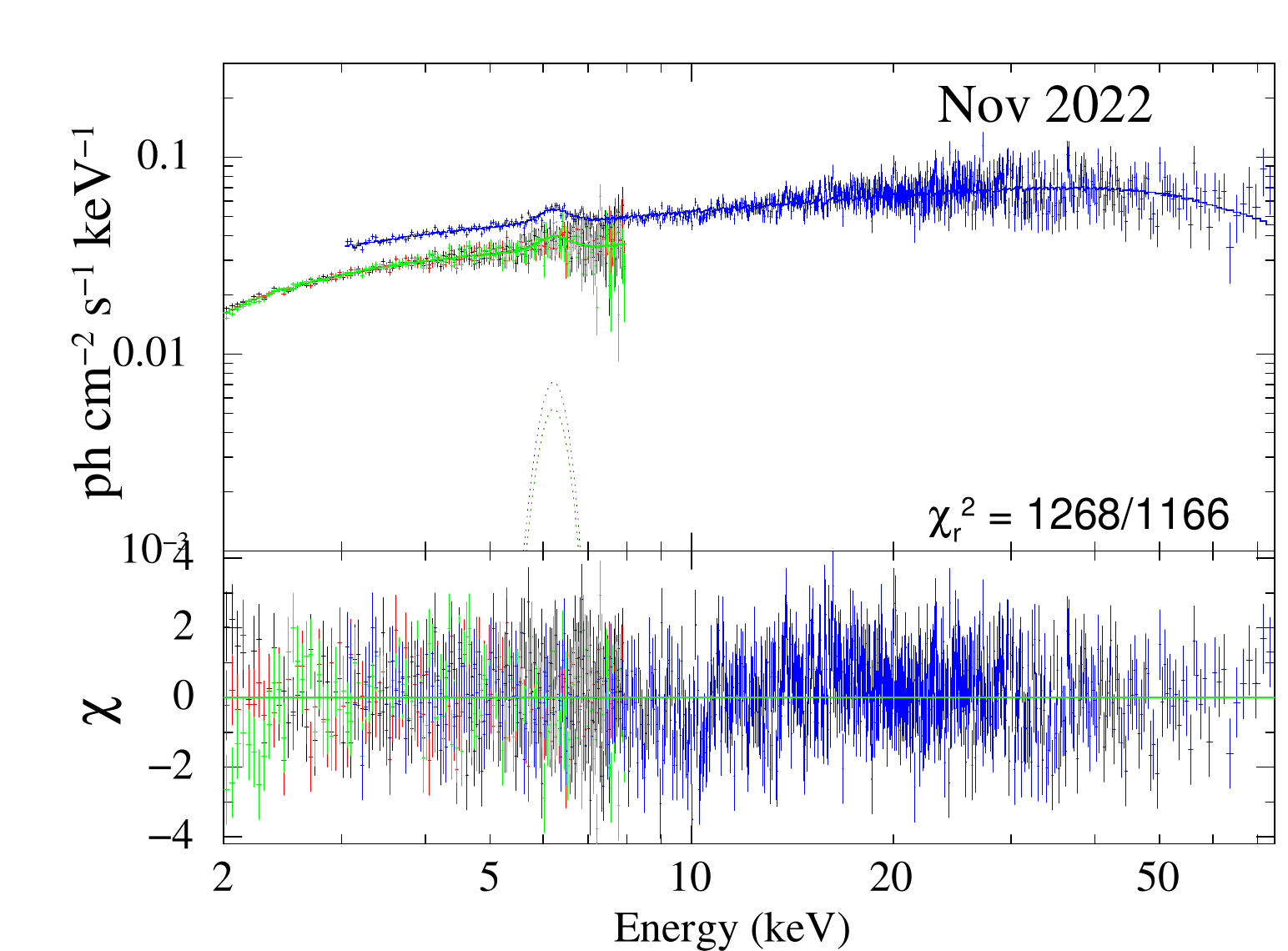}}
\caption{Joint spectral modelling of {\it IXPE} and {\it NuSTAR} data fitted using \jetcaf\,model in the energy band $2-79$ keV. The left and right panels are for the epochs (I1+N1) and (I2+N2). The Fe K$\alpha$ line $\sim 6.3$ is fitted using \gauss\, model as shown by dashed lines. The black, red, and green points correspond to {\it IXPE} data for all three DUs and the blue points correspond to {\it NuSTAR} data. See the text for details.
} 
\label{fig:SpecFit}
\end{figure*}

In Fig. \ref{fig:SpecFit}, we show the model-fitted spectra of MCG-5-23-16 for the joint {\it IXPE} and {\it NuSTAR} observations. The Fe K$\alpha$ line fitted well with a Gaussian line width of $\sim 0.3$ keV for all observations. The Compton hump part of the spectra fitted with \jetcaf\, model is taken into account by the scattering of corona photons by the diverging outflow. The disk mass accretion rate in \jetcaf\, model varied minimally, whereas the halo accretion rate was nearly constant $\sim 0.4$ $\dot m_{\rm Edd}$ and the size of the corona changed significantly from $\sim 28$ to $40 r_s$. The model fit to all the epochs gives nearly a constant mass within errors of the central SMBH $\sim 2\times 10^7$ M$_\odot$, consistent with previous estimations reported in the literature. However, if we freeze the mass to a known value of $2\times 10^7$ M$_\odot$ from the literature, it will provide consistent fits. That will also reduce the errors noticeably in $\dot m_d$ and $\dot m_h$ and marginally in other parameters. For the joint epochs data (I1+N1 and I2+N2) fits, the $N_H$ for the host galaxy absorption required in \ztbabs\, model is $2.0-2.2\times 10^{22}$ cm$^{-2}$, while earlier {\it NuSTAR} epochs (N3-N7) required $N_H \sim 1.2-1.5 \times 10^{22}$ cm$^{-2}$. The high value of the $f_{\rm col}$ parameter implies that the outflow is not well collimated and the rate is high. The estimated outflow solid angle is $\sim 0.5\pi$. The \jetcaf\, model fitted parameters are summarized in Table \ref{tab:parsJetcaf}. 

\begin{table*}
    \centering
    \small
    \caption{Joint {\it IXPE} and {\it NuSTAR} data fitted model parameters are provided in this table. Here, $M_{\rm BH}$, $\dot m_d$, $\dot m_h$, $X_s$, $R$, and $f_{\rm col}$ are the mass of the BH, disk and halo mass accretion rates, location of the shock or size of the corona, the shock compression ratio, and the outflow collimation factor respectively in \jetcaf\,model. N$_H$ is the hydrogen column density in \ztbabs\,model. All data sets required \gauss\, model for the Fe K$\alpha$ line $\sim 6.3$ keV of width $\sigma_g$ given in the table.}
\begin{tabular}{ccccccccccc}
\hline
Model&\multicolumn{6}{c}{\jetcaf}&\ztbabs&\multicolumn{1}{c}{\gauss}\\
\hline
Epoch  &$M_{\rm BH}$ &$\dot m_{\rm d}$ & $\dot m_{\rm h}$ & $X_{\rm s}$ & R &$f_{\rm col}$&$N_H$&$\sigma_g$ &$\chi^2/dof$ \\
	          & $(10^7 M_\odot)$&$(10^{-2}\dot m_{\rm Edd})$&$(\dot m_{\rm Edd})$&$(r_{\rm s})$& & & $(10^{22}$ cm$^{-2})$&(keV) \\
\hline
     I1+N1   &$2.0\pm0.3$&$1.3\pm0.2$&$0.41\pm0.03$&$38.3\pm4.4$&$4.91\pm0.51$&$0.53\pm0.07$ &$2.21\pm0.25$&$0.30\pm0.03$&1329/1180 \\
     I2+N2   &$2.1\pm0.2$&$0.9\pm0.1$&$0.40\pm0.03$&$39.4\pm3.4$&$4.36\pm0.23$&$0.49\pm0.04$ &$2.04\pm0.39$&$0.31\pm0.04$&1268/1166 \\
     N3   &$2.1\pm0.2$&$1.2\pm0.2$&$0.40\pm0.02$&$28.2\pm3.1$&$5.24\pm0.38$&$0.45\pm0.04$ &$1.37\pm0.11$&$0.27\pm0.02$&1102/989 \\
     N4   &$1.9\pm0.2$&$1.3\pm0.1$&$0.41\pm0.03$&$34.9\pm3.9$&$5.19\pm0.32$&$0.49\pm0.05$ &$1.22\pm0.09$&$0.31\pm0.03$&864/768 \\
     N5   &$2.0\pm0.1$&$1.3\pm0.1$&$0.41\pm0.02$&$29.9\pm2.1$&$5.28\pm0.42$&$0.47\pm0.03$ &$1.49\pm0.15$&$0.31\pm0.02$&1029/928 \\
     N6   &$2.1\pm0.1$&$1.1\pm0.1$&$0.41\pm0.04$&$39.8\pm4.6$&$4.98\pm0.37$&$0.54\pm0.03$ &$1.27\pm0.07$&$0.35\pm0.03$&983/930 \\ 
     N7   &$2.0\pm0.1$&$1.3\pm0.1$&$0.42\pm0.02$&$37.9\pm2.8$&$5.06\pm0.30$&$0.53\pm0.04$ &$1.14\pm0.15$&$0.27\pm0.02$&609/570 \\
     \hline     
    \end{tabular} 
    \label{tab:parsJetcaf}
\end{table*}

From the \jetcaf\,model fitted parameters, we have estimated different physical quantities to understand the variation of polarization with corona geometry and spectral flux. The height of the corona is basically the height of the shock, as the shock is the boundary layer of the corona. The height of the shock ($h_{\rm shk}$) is estimated using  \autocite{Debnathetal2014}, $h_{\rm shk}=[\gamma (R-1) X_s^2/R^2]^{1/2},$
where, $\gamma$ is the adiabatic index of the flow, which is considered here as 5/3. The estimated $h_{\rm shk}$ varies in a range between $\sim 14 - 21$ $r_s$.

The mass outflow rate ($\dot m_{\rm out}$) is derived for an isothermal base of the jet \autocite{Chakrabarti1999}, which is given by,
\begin{equation}
\dot m_{\rm out}=\dot m_{\rm in} f_{\rm col}f_0^{3/2}\frac{R}{4} \exp\left[\frac{3}{2} -f_0\right],
\end{equation}
where, $f_{\rm 0}=\frac{R^2}{R-1}$ and $\dot m_{\rm in} (=R\dot m_h+\dot m_d)$ is the total mass inflow rate. As the matter gets compressed at the shocked region by a factor of $R$, it changes the optical depth of the corona \autocite[see][]{ChakrabartiTitarchuk1995,MondalEtal2014Ap&SS.353..223M}. The estimated $\dot m_{\rm out}$ varies between $0.14\pm0.03-0.20\pm0.03$ $\dot M_{\rm Edd}$ for this source, which is significant as was reported earlier in the literature \autocite[see,][]{BraitoEtal2007ApJ...670..978B}. According to this model, as mentioned earlier, both corona and base of the jet behave as a Comptonizing medium. The radiation scattered by the corona or wind may exhibit significant polarization. The unpolarized radiation from the central source scattered by the corona of radius $X_s$ at height $h_{\rm shk}$ above the midplane of the disk will acquire a net fractional polarization \autocite{BegelmaMcKee1983ApJ...271...89B},

\begin{equation}\label{eq:2}
    \Pi_{\rm sc}=\frac{\sin^2i[1-2(h_{\rm shk}/X_{s})^2]}{2[1+2(h_{\rm shk}/X_{s})^2]+[1-2(h_{\rm shk}/X_{s})^2]\sin^2i},
\end{equation}
where $i$ is the disk inclination angle. As the above estimation uses only Thomson scattering as an approximation of Compton similar to \jetcaf\,model, thereby, the scattering conserves photon energy.  Moreover, the Thomson scattering has no frequency dependence so it would preserve the incident spectrum for any energy bands including {\it IXPE} coverage. Therefore, for a given or well-estimated corona properties and disk inclination, one can estimate PD or vice-versa using Eq. \ref{eq:2}. Thereby, to shed light on the estimation of $i$ from the polarization measurement, we have calculated it from the observed upper limit of the PD in \cite{MarinucciEtal2022MNRAS.516.5907M,TagliacozzoEtal2023MNRAS.525.4735T} and the broadband spectroscopic data fitted \jetcaf\,model parameters. Our estimated $i$ ranges between $31^\circ - 35^\circ$ during epochs I2 and I1 for given observed PDs. The rest of the {\it NuSTAR} epochs do not have polarization data, thereby the measurement of observed PD values. However, we can predict the expected PDs by taking an average value of $i$ (which is $33^\circ$) from I1 and I2 epochs and using it in Eq. 2. Our estimation is a bit lower compared to the estimations reported in the literature, however, falls in the range estimated using reflection models (see the next section). Therefore, we have further estimated the PD values for $i=45^\circ$ as a comparison with the literature. All estimated PD values are given in Table \ref{table:PolCorona}. The spectral flux given in the table is calculated using {\it flux err} command in {\it XSPEC} for the energy band $2-79$ keV.

\begin{table*}
\centering
\caption{\label{table:PolCorona} Estimated and observed polarization and corona properties of MCG-5-23-16. The $3-79$ keV band flux is in units of $10^{-10}$ erg cm$^{-2}$ s$^{-1}$. Here, $^{a,b}$ denotes the references to \cite{MarinucciEtal2022MNRAS.516.5907M} and \cite{TagliacozzoEtal2023MNRAS.525.4735T}. In the first two rows, $^c$ denotes the estimation of $i$ from the observed upper limits of PD, while the expected PDs in column 5 are calculated using \jetcaf\,model geometry and for $i=33^\circ$, which is an average of the estimates in the first two epochs ($35^\circ$ and $31^\circ$ respectively). The column 8 denotes the estimations of PD for a given angle of $45^\circ$, which is considered from the literature. The last two columns stand for the same estimation of PD for a given $i$, which is obtained from \relxil\,model and using the geometry from \jetcaf\,model.}
\begin{tabular}{cccccccccccc}
\hline
  Epoch &$\dot m_{\rm out}$  & $h_{\rm shk}$ &$F_{\rm 3-79 keV}$ &\multicolumn{2}{c}{PD(\%)}  &$i^\circ$ &PD(\%)&PD(\%)&$i^\circ$ \\
           & $(\dot m_{\rm Edd})$ & ($r_s$)&  &Estm.  & Obs.$^{a,b}$  &\jetcaf&Estim. &Estim.&\relxil   \\
\hline
I1+N1&$0.19\pm0.04$&$19.9\pm2.5$&$2.75\pm0.02$&$4.2$&$\leq4.7$&$35^{c}$&6.9&4.9&35.9\\
I2+N2&$0.20\pm0.03$&$21.4\pm1.9$&$2.84\pm0.02$&$3.7$&$\leq3.3$&$31^{c}$ &6.1&3.4&31.4\\
\cline{1-7}
 N3&$0.14\pm0.03$&$14.3\pm1.7$&$2.66\pm0.03$&$4.5$&-&33 &7.4&7.4&44.8\\
 N4&$0.16\pm0.02$&$17.8\pm2.1$&$2.53\pm0.01$&$4.5$&-&33 &7.3&5.8&38.6\\
 N5&$0.15\pm0.02$&$15.1\pm1.2$&$2.30\pm0.02$&$4.6$&-&33 &7.5&6.0&39.1\\
 N6&$0.19\pm0.03$&$20.6\pm2.5$&$3.09\pm0.02$&$4.3$&-&33 &7.0&5.9&39.9\\
 N7&$0.19\pm0.02$&$19.5\pm1.6$&$2.87\pm0.02$&$4.4$&-&33 &7.2&5.7&38.8\\
\hline
\end{tabular}
\end{table*}

\begin{figure*}
	\centering 
	\includegraphics[width=0.3\textwidth, angle=0]{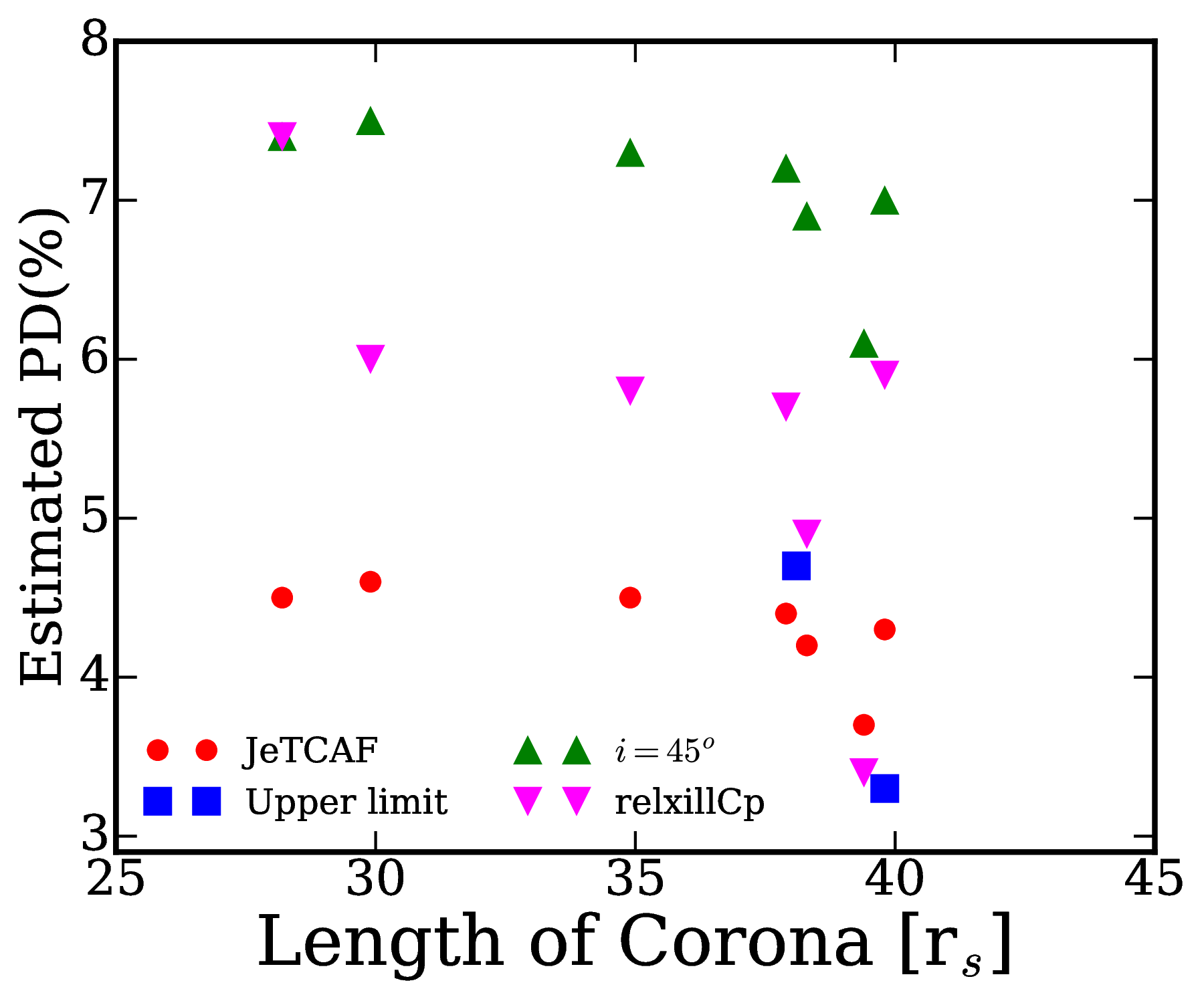}
    \includegraphics[width=0.3\textwidth, angle=0]{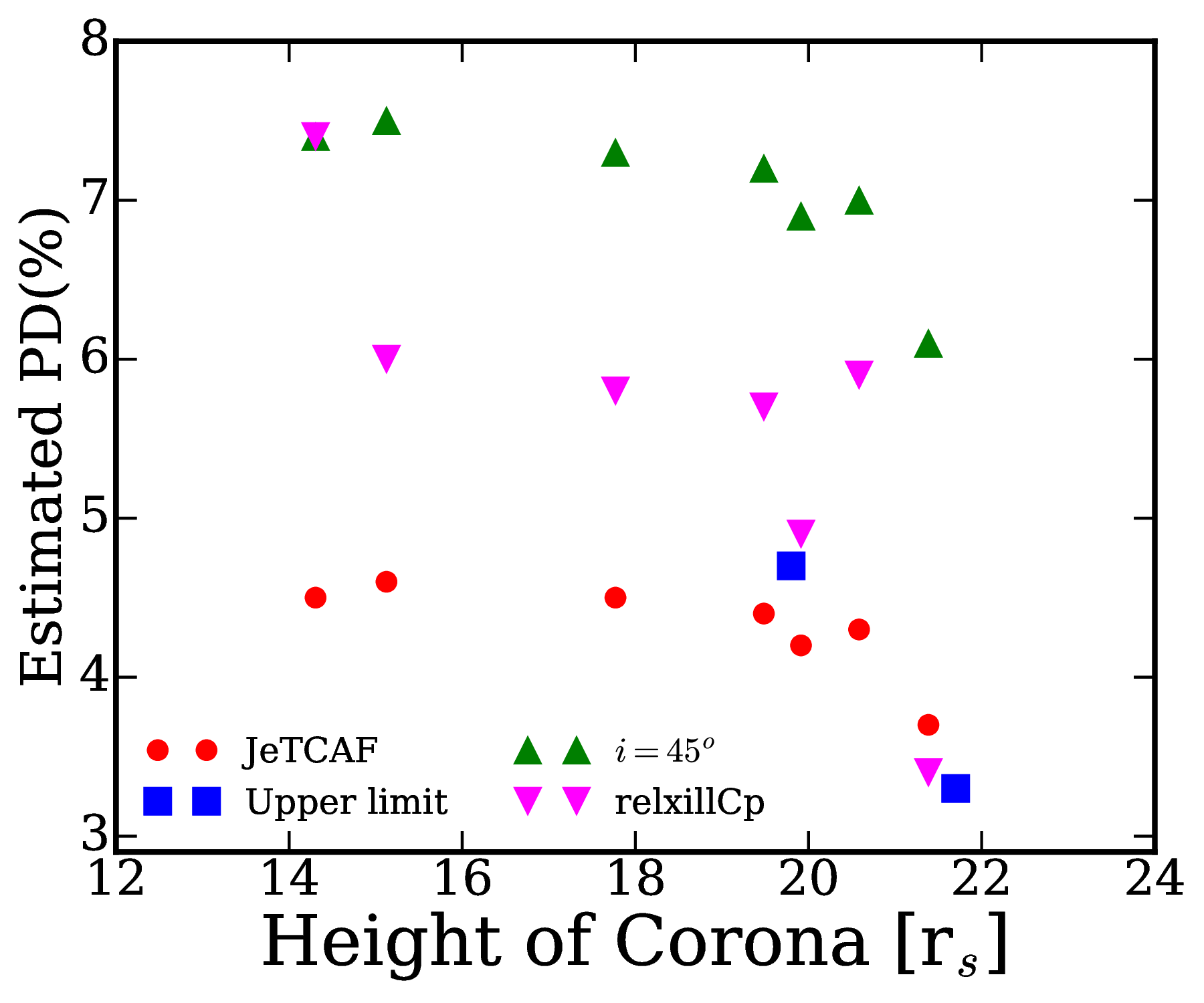}
    \includegraphics[width=0.3\textwidth, angle=0]{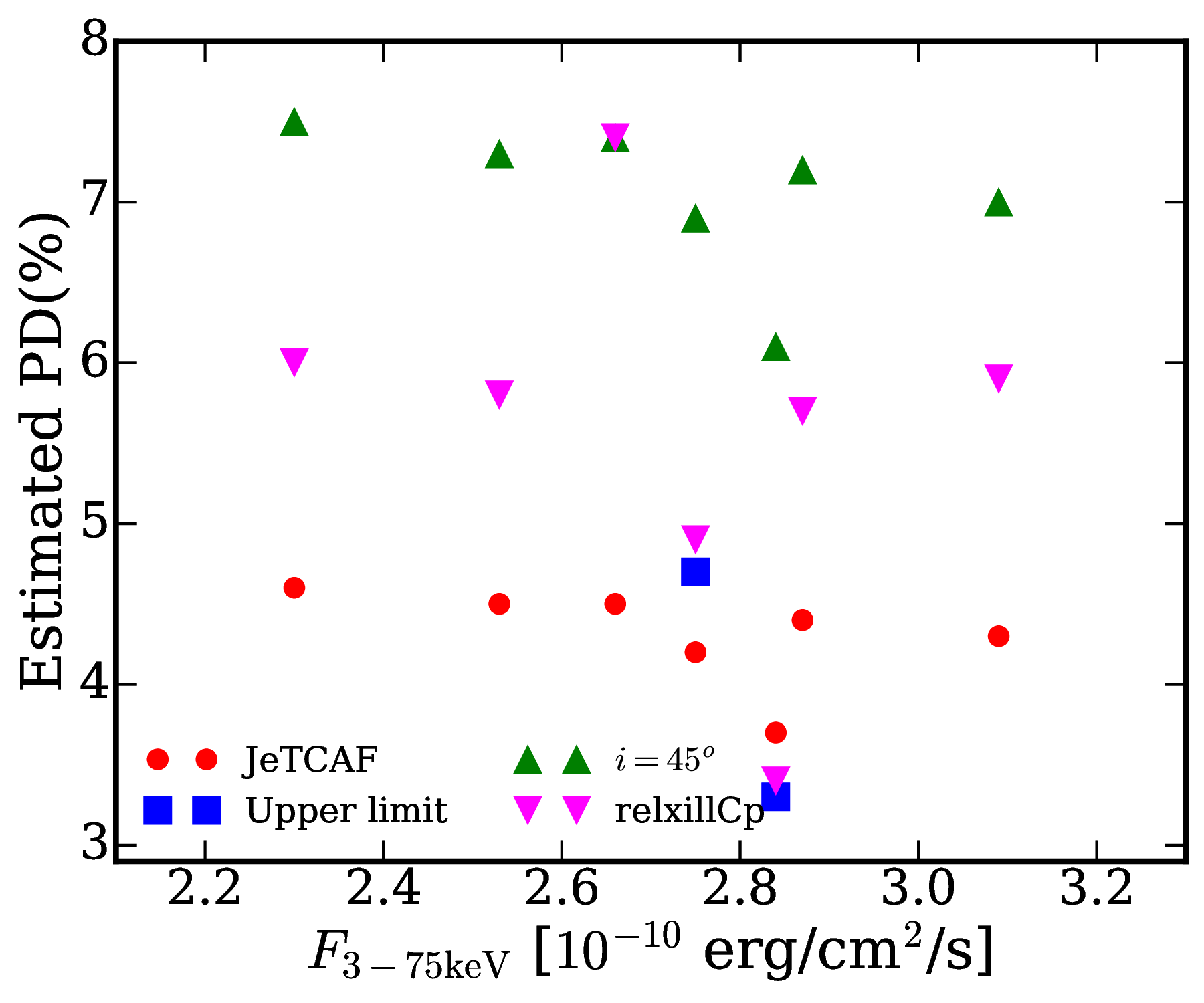}
	\caption{Dependence of PD with the geometry of the Comptonizing corona in \jetcaf\,model. The left and middle panels show the PD variation with the radial distance and height of the corona, while the right panel shows the PD variation with the observed flux in the $3-79$ keV band. The red circles show the predicted PD, estimated using spectral model fitted accretion parameters, and the blue squares show the observed upper limit in PD measurement. The magenta upside down triangles and green triangles show the PD estimations from \relxil\,model fitted disk inclination and for a given inclination of $45^\circ$.} 
	\label{fig:PolGeo}%
\end{figure*}

The left panel of Fig. \ref{fig:PolGeo} shows the variation of PD with the radial distance of the corona. The red circles are the PD values estimated using the model fitted $h_{\rm shk}$ and $X_s$ parameters for a given $i=33^\circ$, which is the average of disk inclination angles estimated from the upper limit of the observed PD values reported in \cite{TagliacozzoEtal2023MNRAS.525.4735T}. For a corona of $<35$ r$_s$, the PD was nearly constant and it decreased when the corona size increased. A similar profile has been observed (middle panel of Fig. \ref{fig:PolGeo}) for the height of the corona, which is quite expected as a bigger corona indicates a hotter electron cloud and more puffed up in the vertical direction. Our estimated PD variation with corona geometry is in accord with the simulation results performed by \cite{SchnittmanKrolik2010ApJ...712..908S} for a spherical geometry of the corona. The right panel of Fig. \ref{fig:PolGeo} shows the PD variation with estimated spectral flux, which shows a negative correlation as well. The variations of PD values estimated for $i=45^\circ$ with the corona properties and model flux are shown by green triangles in all three panels of Fig. \ref{fig:PolGeo}. For the estimation, we used the corona geometry obtained from \jetcaf\,model fits. The estimated PD values are higher by nearly a factor of two than the observed and \jetcaf\,model estimations. A more spherical corona might be required for the same disk inclination to get a lower value of PD which can match with observations.

\subsubsection{Reflection based \pexrav\,and \relxil\,models}
Further, we have used reflection-based \pexrav\, and 
\relxil ~models to fit the joint and individual spectra to constrain different reflection parameters. The models reads in \texttt{XSPEC} as \tbabs*\ztbabs*(\gauss+\pexrav) and 
\tbabs* \ztbabs*\relxil, and a \constant\, component for the first two epochs. For both model fittings, we set the abundances to their Solar value.  First, we fit the data using \pexrav\, model, where the reflection parameter ($R_{\rm ref}$) varied between $0.46\pm0.14-0.69\pm0.22$, the powerlaw photon index $1.78\pm-1.94\pm0.03$. The FeK$\alpha$ line fitted $\sim 6.3$ keV with width $\sigma_g$ as tabulated in Table \ref{tab:parsRefl}. From this model fit, we could constrain the cutoff energy ranges between $130-184$ keV. These values were used during fitting using \relxil\,model. To constrain the disk inclination angle ($i$), we put a range of $0.6<i<0.9$. For some epochs (N3-N5), the $i$ value is pegged at the lower limit, therefore, we could not constrain its value. However, the other epochs returned the inclination angle value between $35^\circ-54^\circ$. The $N_H$ values for the first two epochs are between 2.25 and $2.50\times10^{22}$ cm$^{-2}$, while for the epochs N3-N7 it is between $1.60-1.86\times10^{22}$ cm$^{-2}$, showing a similar trend as obtained in \jetcaf\,model.

\begin{figure*}
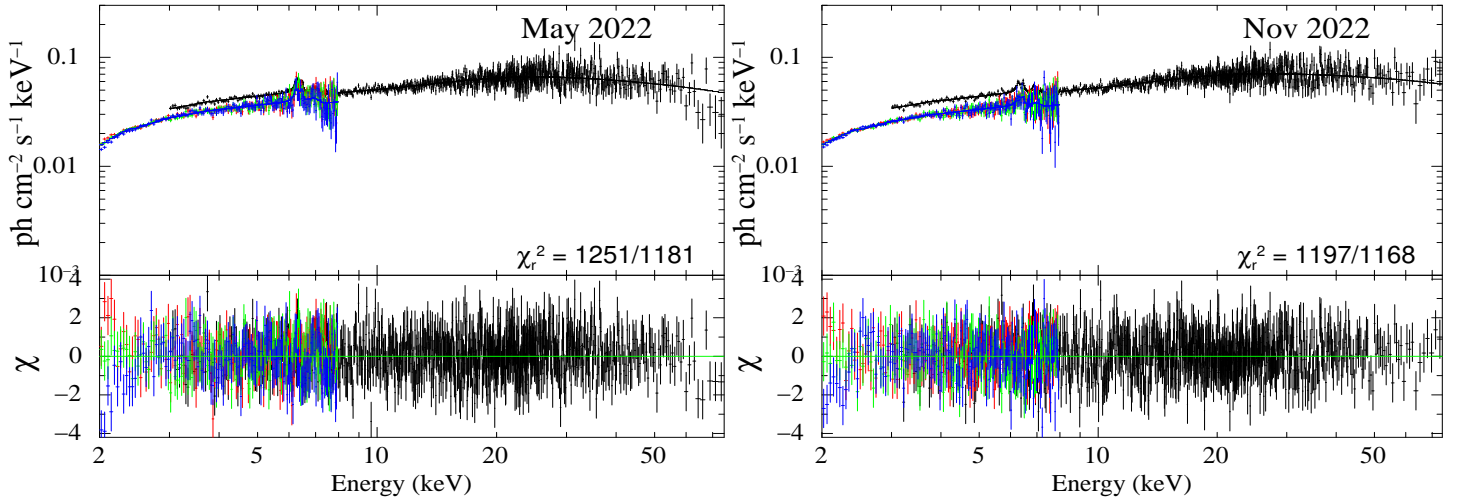

\centering{
\hspace{-1.5cm}
\includegraphics[height=10.0truecm,width=7.0truecm,angle=270]{Figures/IN-may-rel.eps}
\hspace{-0.6cm}
\includegraphics[height=10.0truecm,width=7.0truecm,angle=270]{Figures/IN-nov-rel.eps}}
\caption{ Same as Fig. \ref{fig:SpecFit} but using \relxil\,model. See the text for details. 
} 
\label{fig:RelSpec}
\end{figure*}

\begin{table*}
    \centering
    \caption{\label{tab:parsRefl} Joint {\it IXPE} and {\it NuSTAR} data fitted reflection model parameters are provided in this table. All data sets required \gauss\, model for the Fe K$\alpha$ line $\sim 6.3$ keV of width $\sigma_g$ given in the table during \pexrav\,model fitting. During both \relxil\, and \pexrav\,model fitting both abundances are set to Solar value. Here, $^p$ and $^f$ denote the pegged and frozen values of the parameters.}
\begin{tabular}{ccccccccc}
\hline
& &\multicolumn{7}{c}{Epochs}\\
\cline{3-9}

Models&Parameters&I1+N1&I2+N2&N3&N4&N5&N6&N7\\
\hline
\ztbabs                    &$N_H$ ($10^{22}$ cm$^{-2}$)&$2.50\pm0.17$&$2.25\pm0.15$&$1.60\pm0.23$&$1.58\pm0.34$&$1.86\pm0.26$&$1.82\pm0.24$&$1.86\pm0.28$\\
\gauss                     &$\sigma_g$ (keV)&$0.18\pm0.03$&$0.19\pm0.04$&$0.18\pm0.02$&$0.18\pm0.03$&$0.22\pm0.02$&$0.24\pm0.02$&$0.17\pm0.05$\\			   
\pexrav                   &$\Gamma$&$1.94\pm0.03$&$1.89\pm0.03$&$1.78\pm0.02$&$1.83\pm0.04$&$1.81\pm0.03$&$1.86\pm0.03$&$1.88\pm0.06$\\
			   &$E_c$ (keV)&$180.5\pm30.6$&$184.3\pm33$&$135.5\pm13.9$&$152.6\pm27.1$&$145.3\pm17.8$&$129.6\pm13.9$&$184.3\pm31.6$\\
			   &$R_{\rm ref}$&$0.58\pm0.22$&$0.46\pm0.21$&$0.46\pm0.14$&$0.54\pm0.23$&$0.58\pm0.17$&$0.56\pm0.16$&$0.69\pm0.08$\\
			   &$\cos(i)$&$0.81\pm0.18$&$0.82\pm0.23$&$0.6^p$&$0.6^p$&$0.6^p$&$0.74\pm0.26$&$0.69\pm0.22$\\
Fit Statistics&$\chi^2/dof$&1260/1182&1198/1169&1020/987&830/869&945/926&877/929&567/571\\ 
\hline 
\ztbabs       &$N_H$($10^{22}$ cm$^{-2}$)&$2.91\pm0.16$&$2.34\pm0.14$&$1.93\pm0.19$&$1.98\pm0.32$&$1.95\pm0.40$&$1.69\pm0.16$&$1.78\pm0.28$\\

\relxil      		   &$i$ $(^\circ)$&$35.9\pm10.7$&$31.4\pm7.9$&$44.8\pm7.3$&$38.6\pm8.8$&$39.1\pm7.4$&$39.9\pm7.5$&$38.8\pm9.4$\\
             		   &$q_2$&$0.1^p$&$0.1^p$&$0.1^p$&$0.1^p$&$0.1^p$&$1.87\pm0.40$&$0.1^p$\\
			   &$\Gamma$&$1.99\pm0.02$&$1.93\pm0.02$&$1.90\pm0.01$&$1.95\pm0.01$&$1.94\pm0.01$&$1.95\pm0.01$&$1.94\pm0.02$\\
			   &$\log\xi$ &$0.93\pm0.34$&$1.28\pm0.18$&$1.14\pm0.10$&$0.57\pm0.44$&$0.70\pm0.33$&$1.22\pm0.18$&$1.56\pm0.38$\\
			   &$N(\text{cm}^{-3})$&$18.05\pm0.45$&$18.19\pm0.29$&$19.57\pm0.29$&$18.07\pm0.20$&$18.16\pm0.28$&$19.41\pm0.30$&$18.27\pm0.57$\\
			   &$kT_e^f$ (keV)&$181$&$184$&$136$&$153$&$145$&$130$&$184$\\
			   &$R_{\rm ref}$&$0.61\pm0.01$&$0.48\pm0.08$&$0.51\pm0.09$&$0.63\pm0.08$&$0.75\pm0.13$&$0.60\pm0.07$&$0.67\pm0.19$\\
Fit Statistics&$\chi^2/dof$  &1251/1181&1197/1168&1022/988&822/870&943/927&896/930&569/572\\     
\hline         
\end{tabular} 
\end{table*}


The relativistic reflection model (\texttt{relxill}) has several variants, including \relxil, which we have used for our purposes. This model takes into account the relativistic disk reflection component and \texttt{NTHCOMP} \citep[][]{ZdziarskiEtal1996MNRAS.283..193Z,ZyckiEtal1999MNRAS.309..561Z} thermal Comptonization continuum component. The \relxil\,model has fourteen parameters out of which two are abundances which we set to their Solar values. If we keep all parameters free, that will give more degenerate values of the parameters and most of the parameters can not be well-constrained. Therefore, before fitting the data we make an educated choice of some of the parameters as was done by several authors in the literature \autocite[][and references therein]{BaruaEtal2023ApJ...958...46B,SerafinelliEatal2023MNRAS.526.3540S}. From the rest of the parameters, we choose non-rotating BH at present, as there is no reporting of the spin parameter for this system. We tried to fit the data by keeping the spin parameter free, and it was always taking the upper hard limit of 0.998. Such a high spin can not be explained from the present data sets as the source evidenced a stable accretion disk being truncated at a larger radius \autocite[see][for more details]{SerafinelliEatal2023MNRAS.526.3540S}. This task prompted us to set $a$ and $R_{\rm in}$ parameters to 0 and 6 $r_g$ in this work. The outer edge of the disk, $R_{\rm out}$ is set to 1000 $r_g$. The parameter $R_{br}$ determines a break radius that separates two regimes with different emissivity profiles and is set to 38 $r_g$, following the previously reported results in the literature and \jetcaf\,model fitted results.  We fixed the first emissivity index ({\it index1}) to 3.0 and put a range for the {\it index2} ($q_2$) between 0.1-4. We set $kT_e$ fixed to values obtained from \pexrav\,model fit. The remaining six parameters $i$, $\Gamma$, ionization index ($\log\xi$), density ($N$), and $R_{\rm ref}$ are set free to vary.

Fig. \ref{fig:RelSpec} shows the \relxil\,model fitted the joint spectra for the epochs May (left panel) and November (right panel). The model fits the data well and the residual is shown in the bottom panels. Using the \relxil\,model, we could well-constrain the disk inclination $i$ between $31.4\pm7.9-44.8\pm7.3$ and the $\Gamma$ was nearly constant across all epochs with value between $1.90\pm0.01-1.99\pm0.02$. The estimated value of $i$ from \jetcaf\,model as given in Table \ref{table:PolCorona} falls within the range of values obtained from \relxil\,model. This range of inclination angle is also in accord with the estimation in the literature \autocite{MarinucciEtal2022MNRAS.516.5907M,SerafinelliEatal2023MNRAS.526.3540S}. The emissivity index ($q_2$) is pegged at 0.1 for all epochs except the epoch N6, for which we obtained a value $1.87\pm0.40$. Our model fits yield ionization parameter $\log\xi \leq 1.6$, which infers that the distant reflecting material is not highly ionized. The obtained density of the matter ($N$) at $R_{in}$ is high, varies between $10^{18-19.6}$ cm$^{-3}$, and the reflection fraction $R_{ref}$ between $0.5-0.8$, which is relatively low. Both reflection model fitted $R_{ref}$ are in close agreement with each other. All reflection model parameters are given in Table \ref{tab:parsRefl}. We have further estimated the PD values using \relxil\,model fitted $i$ values. The results are summarized in Table \ref{table:PolCorona}. The variations of PD estimated using \relxil\,model fitted $i$ values are shown by upside-down magenta triangles in all three panels of Fig. \ref{fig:PolGeo}. It is noticeable that these PD values fall in between the observed and for $i=45^\circ$ estimations.

to appear in While using two reflection-based models to the same set of data, both models fit the data equally well provided some marginal changes in $\chi^2$ values. However, from the \pexrav\,model we could not constrain the $i$ parameter, provided a broad range. The \relxil\, could constrain the $i$ values which are in agreement with the literature, and we could use them to compare the observed PD measurement. Since the \pexrav\,model could constrain the $E_{cut}$, we could use them in \relxil\,model. Other common parameters ($\Gamma$ and $R_{ref}$) obtained from the fit in both models are quite similar. We note that, the \relxil\,model takes into account the ionized reflection, ionization, and relativistic effects, could constrain the required parameter and, therefore more favorable for the present study, compared to the \pexrav\,model.

\section{Discussions and Conclusions}
In this work, we have performed the spectro-polarimetric analysis of the Seyfert galaxy MCG-5-23-16 to understand the polarization properties and the X-ray emitting corona using two epochs of data from {\it IXPE} and a decade of observations from {\it NuSTAR}. Our model-independent analysis poses the PD with an upper limit of $\leq 3.8\%$, while the model-dependent estimates give PD $1.08\pm0.66\%$ and PA $55.6^\circ\pm19.1^\circ$ using simultaneous data fitting of I1 and I2, which are in accord with the previous studies within error bar \autocite{MarinucciEtal2022MNRAS.516.5907M,TagliacozzoEtal2023MNRAS.525.4735T}.

The joint {\it IXPE} and {\it NuSTAR} broadband spectroscopic data fitted using \jetcaf\,model gives the estimate of disk mass accretion rate to the central SMBH which is $\leq 0.013$ $\dot m_{\rm Edd}$, while the halo accretion rate is $\leq 0.4$ $\dot m_{\rm Edd}$. The location of the shock changed from 28 to 40 $r_s$ and the shock compression ratio varied between $4.4-5.3$. Relatively high values of jet/outflow collimation factor indicate that the outflow is not well-collimated. The previous studies reported that the source showed significant mass outflow, here we estimated the mass outflow rate $\sim 0.14-0.20$ $\dot m_{\rm Edd}$ from the spectral model fitted parameters. Since mass outflow carries thermal energy from the hot corona, its geometry changes \autocite{Chakrabarti1999,MondalEtal2014Ap&SS.353..223M}. Therefore, the final best-fit corona geometry which is used to estimate the expected PD has the effect of mass outflows. Hence, studying polarization using the disk-corona-outflows model is crucial to get a consistent picture of a system, which has been done in this work. Since the model directly fits the data using corona geometry and estimates expected $i$ using observed quantities, such estimations might be more favorable.

We have further fitted the data using \pexrav\,and \relxil\,models to constrain the reflection properties and disk inclination ($i$). We found that the reflection fraction, $R_{\rm ref}<0.8$ which agrees with previous studies, and the photon index $\Gamma$ was also consistent in both models. In \pexrav\,model, we could constrain $i$ for some epochs, and for some epochs fits were trending towards very high inclination. However, using \relxil\,we could constrain $i$ for all epochs which put a range $\sim 31^\circ-45^\circ$, which is in agreement with the previous studies. Therefore, the \relxil\,is a more favorable reflection model for the present data sets compared to \pexrav. Given these values of $i$ from \relxil\,model, we have estimated the expected PD to be 3.4-6.0\%, which is comparatively similar to higher than the observed PD.

The scattering of photons inside the Comptonizing corona or outflow can induce significant polarization as discussed earlier. Therefore, we used it to further estimate the disk inclination for a given corona geometry from \jetcaf\,model, mainly its radial distance and height. Using the upper limit of polarization measurement, we have estimated the disk inclination $\sim 35^\circ$ using observations in May 2022 (I1+N1) and $\sim 31^\circ$ in November 2022 (I2+N2). After taking an average of these two estimates, we further use it to determine the expected polarization during the epochs N3-N7. The estimated PD decreases with increasing the corona geometry at different epochs (see Fig. \ref{fig:PolGeo}). We note that the expected $i$ estimated using the \jetcaf\,geometry for the simultaneous data sets nicely matches with the \relxil\,model fitted $i$. We have further estimated the expected PDs using $i$ from \relxil\,model and also considering $i$ to be $45^\circ$ are shown in all three panels of Fig. \ref{fig:PolGeo} for a comparative study. These values are above the presently observed upper limit which may suggest that the disk inclination predicted from \jetcaf\, and obtained from \relxil\, model is more reliable. However, if such a high PD is expected, that can be possibly measured in the future with broadband polarimetry using joint {\it IXPE} and {\it XPoSat}, which may provide higher inclination, require a further study. 
Our study not only relates the polarization with the accretion flow properties but also can predict expected polarization. 
The current model prescription can be applied to other Seyfert galaxies to study their X-ray polarization in connection with the accretion-ejection properties, where polarization could be due to reflection \autocite[NGC 4151;][]{GianolliEtal2023MNRAS.523.4468G} or outflowing jet \autocite[IC 4329A;][]{IngramEtal2023MNRAS.525.5437I,PalEtal2023JApA...44...87P}, which will be addressed in elsewhere. Additionally, the current modeling can be applied to {\it XPoSat} observations of this source and other potential candidates in the future.

\begin{acknowledgement}
We thank the referee for making constructive comments and suggestions that improved the quality of the paper. 
SM thanks Keith A. Arnaud for helping during the data analysis. 
RC, VKA, and AN thank GH, SAG, DD, PDMSA, and the Director, URSC for continuous support to carry out the research activity. This research used data products provided by the IXPE Team
(MSFC, SSDC, INAF, and INFN) and distributed with additional
software tools by the High-Energy Astrophysics Science Archive Research Center (HEASARC), at NASA Goddard Space Flight
Center (GSFC). 
This research has made use of the {\it NuSTAR} Data Analysis Software (NuSTARDAS) jointly developed by the ASI Science
Data Center (ASDC, Italy) and the California Institute of Technology (Caltech, USA).
\end{acknowledgement}

\paragraph{Funding Statement}

SM acknowledges Ramanujan Fellowship (\# RJF/2020/000113) by SERB-DST, Govt. of India for this research.

\paragraph{Competing Interests}

None

\paragraph{Data Availability Statement}

Data used in this work are publicly available in NASA's HEASARC archive. The \jetcaf\, model \citep{MondalChakrabarti2021} is incorporated in {\it XSPEC} directly using \texttt{initpackage} command, which is not public yet. However, the preliminary version of the FITS file is in a testing phase and can be available soon upon request or on GitHub\footnote{https://github.com/santanumondal87/JeTCAF-A-package-for-X-ray-spectral-fitting-of-black-holes-across-mass-scale}.


\printbibliography

\end{document}